# In-plane emission of indistinguishable photons generated by an integrated quantum emitter.

Sokratis Kalliakos,[1*] Yarden Brody,[1,2] Andre Schwagmann,[1,2] Anthony J. Bennett,[1] Martin B. Ward,[1] David J. P. Ellis,[1] Joanna Skiba-Szymanska,[1] Ian Farrer,[2] Jonathan P. Griffiths,[2] Geb A. C. Jones,[2] David A. Ritchie,[2] and Andrew J. Shields[1]

[1] *Cambridge Research Laboratory, Toshiba Research Europe Limited, 208 Science Park, Milton Road, Cambridge, CB4 0GZ, United Kingdom*

[2] *Cavendish Laboratory, University of Cambridge, J. J. Thomson Avenue, Cambridge CB3 0HE, United Kingdom.*

We demonstrate the emission of indistinguishable photons along a semiconductor chip originating from carrier recombination in an InAs quantum dot. The emitter is integrated in the waveguiding region of a photonic crystal structure, allowing for on-chip light propagation. We perform a Hong-Ou-Mandel-type of experiment with photons collected from the exit of the waveguide and we observe two-photon interference under continuous wave excitation. Our results pave the way for the integration of quantum emitters in advanced photonic quantum circuits.

* email: sokratis.kalliakos@crl.toshiba.co.uk

The use of single photons as flying qubits for linear optics quantum computing has triggered a lot of interest in the scientific community since it was first proposed.[1] Most of the applications in the field of quantum information rely on the two-photon interference effect, where two identical photons impinging on different sides of a 50/50 beamsplitter leave in the same direction. The effect is known as Hong-Ou-Mandel (HOM) interference[2] and relies on the destructive interference of the probability amplitudes of two-photon states. The key for successful HOM interference is the indistinguishability of the two photons in the spectral, temporal and polarization degrees of freedom. Additionally, perfect spatial overlap of the photon wave function is required at the beamsplitter.

Emission of indistinguishable photons has been demonstrated using different emitters such as single molecules,[3] trapped ions,[4] atoms,[5] and semiconductor quantum dots (QDs).[6] As an efficient single-photon source,[7] QDs offer significant advantages for on-chip integration and scaling of the quantum emitter. Both single[8] and indistinguishable[9] photon emission as well as emission of entangled photon pairs[10] has been achieved with devices operating under electrical carrier injection. Their flexibility has been highlighted in recent studies where photon emission from a single QD has been used in HOM-type of experiments in combination with light emitted from different sources.[11,12] They can be efficiently coupled to nanocavities,[13,14] where quantum electrodynamics effects can be utilized to improve device operation. Quantum light devices based on site-controlled quantum dots have been recently demonstrated,[15] highlighting the potential of this system for scalability. Most of the studies were restricted to the out-of-plane light emission properties of the QDs. However, in the emerging field of photonic quantum information technology, it is desirable to integrate the quantum emitter into an on-chip quantum circuit. Recent studies have reported the integration of QDs in photonic circuits, in particular photonic crystal waveguides,[16-21] which extended the use of these emitters as integrated in-plane quantum light sources.

In this letter we demonstrate the indistinguishability of single photons emitted and transmitted along the plane of the semiconductor chip. Single photons are generated by carrier recombination in a QD positioned in the waveguiding region of a photonic crystal waveguide slab. No cavity/QD coupling effects are present. The emitted photon energy lies within the spectral region of the waveguide's propagating mode, allowing for the transfer of quantum light along the chip. For photons emitted from the exit of the waveguide, we observe two-photon interference visibility of $40 \pm 4\%$ limited by the temporal resolution of single-photon detectors. There is excellent agreement between our findings and our theoretical model, whose parameters are determined experimentally.

Our device is composed of a unidirectional photonic crystal W1 slab waveguide. The GaAs photonic crystal slab contains a layer of low-density InAs QDs at the center of the slab that serve as quantum emitters. The device is fabricated by standard electron-beam lithography with a combination of wet etching (for the removal of the sacrificial AlGaAs layer) and dry etching techniques.[18] During the experiments the sample is kept in a cryostat at $T = 5$ K. We probe the light emission into the waveguide by optically exciting the QD with a continuous wave laser ($\lambda = 633$ nm) from the top of the slab while collecting from the waveguide's exit.[18] Previous studies of similar devices with identical design characteristics have shown single-photon injection into the propagating mode of the waveguide with efficiency over 24%.[18]

The in-plane spectrum from the photonic crystal waveguide device with lattice constant $a$ = 229 nm is shown in Fig. 1 (b). At low excitation power, the spectrum is dominated by an emission line from a QD situated in the waveguiding region. It is not clear if it is a neutral or charged exciton state. A polarizer was used to analyze the polarization properties. We find that the emission is highly polarized with the electric field vector pointing along the plane of the slab and perpendicular to the direction of the waveguide (*y*-axis).[18] For comparison, the spectrum recorded with the polarizer's axis set perpendicular to the slab's plane (*z*-axis) is shown at the inset of Fig. 1 (b), where no emission is observed. Considering the fact that the excitonic state has no dipole moment out of the plane of the slab, *z*-polarized QD emission should only occur because of scattering effects. On the other hand, the transmission of *y*-polarized light along the waveguide occurs because of the propagating modes of the waveguide. Fig. 1 (c) shows the band structure of a photonic crystal waveguide, which has been calculated for the case of an infinite waveguide (as shown in Fig. 1 (a)) using the plane-wave expansion method. The emission energy of the QD (dashed line) is resonant with the waveguide's low-energy propagation mode (blue line), which is polarized along the *y*-axis. The QD emission is blue-shifted with respect to the slow-light mode (at ~ 0.244 $a/\lambda$), which does not appear in the spectrum. In other words, light emitted by the carrier recombination in the QD is injected in the polarized propagating mode and transmitted along the waveguide.

We analysed the photon statistics of the in-plane emission of the QD using a Hanbury-Brown and Twiss set-up. Silicon avalanche photodiodes operating in the Geiger mode were used as detectors. Photon coincidences give a direct measure of the second-order correlation function $g^{(2)}(\tau) = \langle I(t)I(t+\tau)\rangle / \langle I(t)\rangle\langle I(t)\rangle$ where *I(t)* is the photon intensity at time *t*. $g^{(2)}(\tau)$ of the emission line shown in Fig. 1 (b) as a function of time delay $\tau$ is shown in Fig. 2 (a). Strong suppression of multi-photon events is observed, with $g^{(2)}(0) = 0.21 \pm 0.04$ extracted using a least squares fitting procedure (red line). This value is limited by the response function of our detectors, which had a temporal resolution of ~ 440 ps. After deconvolution, the multiphoton emission probability drops to zero.

For applications in linear optical quantum computational schemes that rely on two-photon interference, the coherence properties of single photons are essential for the desired functionalities. Since InAs QDs are embedded in a semiconductor matrix, their interaction with phonons and localized charges in the surrounding region is strong, causing unavoidable dephasing.[22,23] We use a Michelson interferometer to measure the coherence time of the single photons emitted in-plane. A simplified illustration of the set-up is shown in Fig. 2 (b) We record the interference fringes by performing a scan with a short piezo-actuated stage at various optical path differences between the two arms of the interferometer (the second arm is controlled by a long translational stage). The path mismatch between the arms is translated into time delay. The visibility of interference fringes at time delay $\tau$ is shown in Fig. 2 (c). An exponential fitting procedure (red line) allows us to extract the single-photon coherence time value $\tau_c = 153 \pm 12$ ps.

We carried out two-photon interference measurements using a polarization-maintaining fiber-based Mach-Zehnder interferometer depicted in Fig. 3 (a). Light emission collected from the exit of the photonic crystal waveguide is dispersed by a transmission grating. Single photons from the QD enter at one input of a 50/50 fiber-based beamsplitter. As already shown, the QD emission is linearly polarized and a half-wave plate is used to perfectly align the polarized photons with the fiber's birefringence axis. The first fiber coupler (splitter) splits the stream of photons to follow two paths, one with a delay line of $\Delta\tau$ = 20 ns and another with an in-line polarization controller. Photons enter the two paths with the same polarization and the controller is used to rotate the polarization state to orthogonal or parallel, rendering the paths distinguishable or indistinguishable, respectively. Photons that travel in indistinguishable arms will destructively interfere at the second 50/50 beamsplitter, provided that the photon arrival times fall within the coherence time $\tau_c$. This will cause photon bunching and a suppression of coincidence events recorded by the detectors. The delay line is chosen to be much longer than the coherence time, assuring no occurrence of lower-order interference.

Second-order correlation measurements for the case of interfering orthogonal and parallel-polarized photons are shown in Fig. 3 (b) and (c) respectively. As with all the measurements presented here, the quantum dot was excited well below saturation in order to achieve low multi-photon emission probabilities and increased coherence times. On the other hand, the excitation power was kept at a level where the count rates at the single photon detectors were high enough (~ 30 kHz) to assure the completion of the correlation measurements within reasonable time. The experiments were performed using superconducting single-photon detectors (SSPDs) with temporal resolution of 140 ps. For a perfect single-photon source, antibunched correlations occur in the case of orthogonally-polarized (distinguishable) 50% of the time. The experimentally attained value is $g_{\text{orth}}^{(2)}(0) = 0.47 \pm 0.04$ (Fig. 3 (b)), remarkably close to the expected value for perfectly antibunched independent emissions. In the case of parallel-polarized photons (Fig. 3 (c)), we observe a dip in the second order correlation function at zero time delay below the classical value of 0.5 (indicated with a dashed line). This is a manifestation of the indistinguishability of the single photons emitted by our device. We measure a second-order correlation function $g^{(2)}_{par}(0) = 0.28 \pm 0.02$. Finally, we determine the visibility of the two-photon interference as $V(\tau) = (g^{(2)}_{orth}(\tau) - g^{(2)}_{par}(\tau)) / g^{(2)}_{orth}(\tau)$ plotted in Fig. 3 (d). At zero time delay, the experimental visibility is $V(0) = 0.40 \pm 0.04$. Within a few hundreds of picoseconds of time delay, the interference visibility drops close to zero, which is consistent with the finite single-photon coherence time.

We use a previously developed[24] theoretical model to fit our experimental data. All of the used parameters were determined experimentally. This includes single-photon coherence time ($\tau_c = 153$ ps), transmission-to-reflection intensity coefficients ratio of the fiber splitters (53/47) and single-photon second-order correlation function $g^{(2)}(0)$ and QD emission lifetime (~ zero and 470 ps respectively, after deconvolution with detector's response function). The predicted second-order correlation functions and visibility are plotted as red curves in Figure 3, taking into account the response function of detectors. There is a very good agreement of the model with the experimental results. The main limiting factor for achieving higher visibility is the temporal resolution of our detector. In the ideal case of a detector with unlimited temporal resolution, the interference visibility is expected to reach unity (blue dashed lines in Fig. 3 (d)).

In conclusion, we have demonstrated the in-plane emission and transmission of indistinguishable photons. The quantum emitter is integrated in the waveguide region of a photonic crystal structure, which assures efficient on-chip transfer of quantum light. We experimentally extract a two-photon interference visibility of 0.40 ± 0.04, limited by the temporal resolution of our single-photon detectors. Theoretical analysis based on experimentally extracted parameters offers good agreement with the experiment. A higher visibility can be anticipated using resonant excitation schemes, while cavity quantum electrodynamics may significantly enhance the efficiency of the device.

**Acknowledgements**

This work was partly supported by the Marie Curie Actions within the Seventh Framework Programme for Research of the European Commission, under the Initial Training Network PICQUE, Grant No. 608062.

**Figure Captions**

Figure 1. (Color online) (a) 2-D illustration of a W1 photonic crystal waveguide, with lattice constant $\alpha$. The axes in the bottom left of the panel refer to the in-plane reference system. (b) In-plane photoluminescence spectrum from the photonic crystal device with polarization set along the slab plane and vertical to the waveguide axis ($y$-polarized). The inset shows the same spectrum of orthogonal polarization ($z$-polarized). (c) Photonic crystal waveguide band structure. The blue lines are the $y$-polarized propagation modes. The emission energy from the QD in (a) is shown with the dashed line.

Figure 2. (Color online) (a) Second-order correlation function for the QD emission as a function of time delay. The red line is the product of a least squares fitting procedure. (b) Schematic illustration of the Michelson interferometer. (c) Visibility of single-photon interference fringes as a function of time delay. An exponential fit (red curve) extracts a coherence time of $153 \pm 12$ ps.

Figure 3. (Color online) (a) Illustration of the set-up for the TPI experiment. Single photons emitted from the exit of the waveguide enter the interferometer from the left side. We achieve correct alignment with the birefringence axis of the polarization-maintaining fibre using a half-wave plate (HWP). The Mach-Zehnder interferometer is composed by two 50-50 fiber couplers. One arm contains a 20 ns delay line and the other a polarisation controller. Superconducting single-photon detectors (SSPDs) are used for detection. (b) Second-order correlation function of cross-polarized and (c) co-polarized interfering photons. Red lines are fits of the theoretical model considering the measured temporal resolution of the detectors. Dashed blue lines represent the ideal case of non resolution-limited detection of events. (d) Visibility of the TPI at different time delays under continuous wave excitation.

**Figure 1**

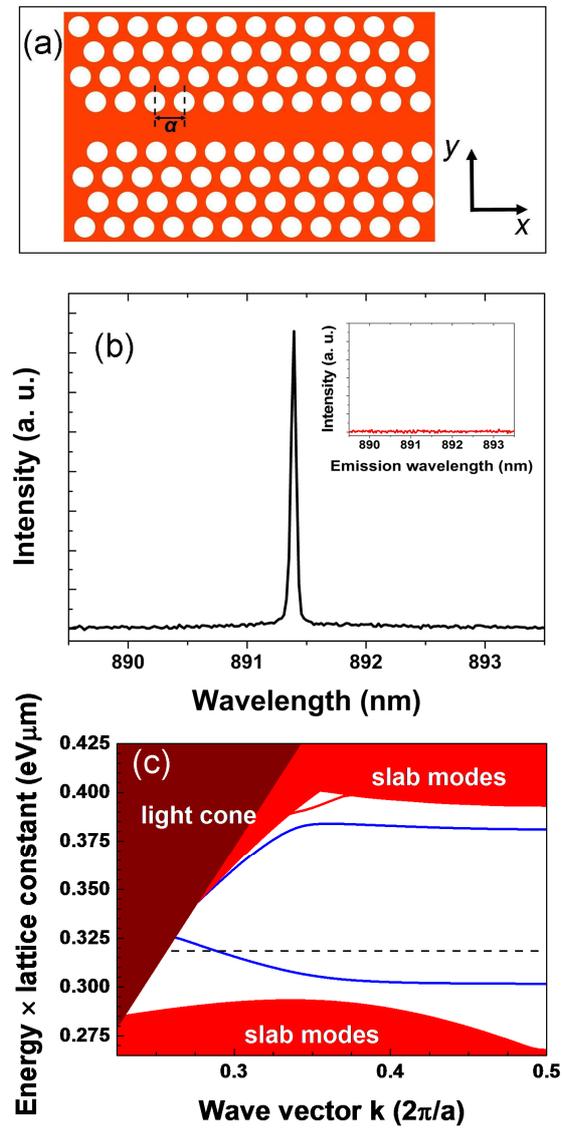

**Figure 2**

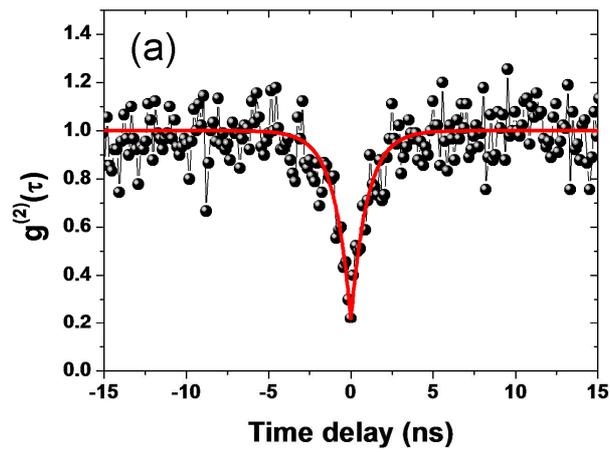

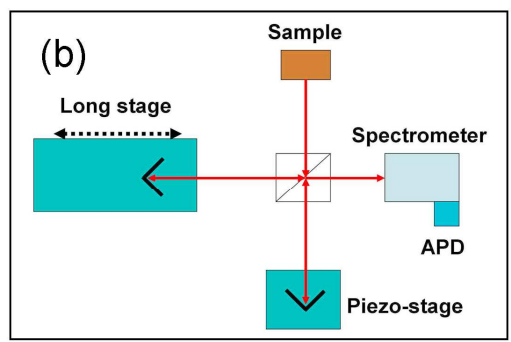

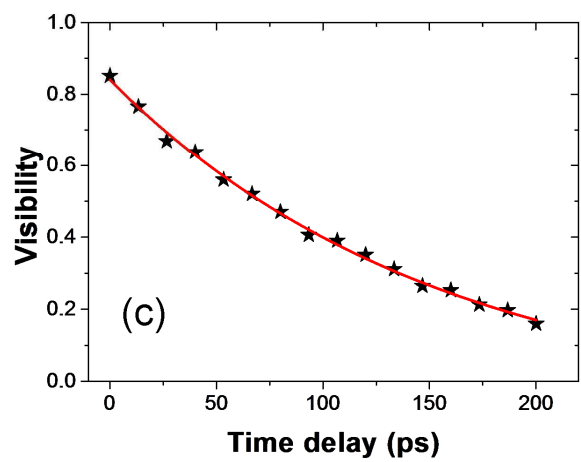

**Figure 3**

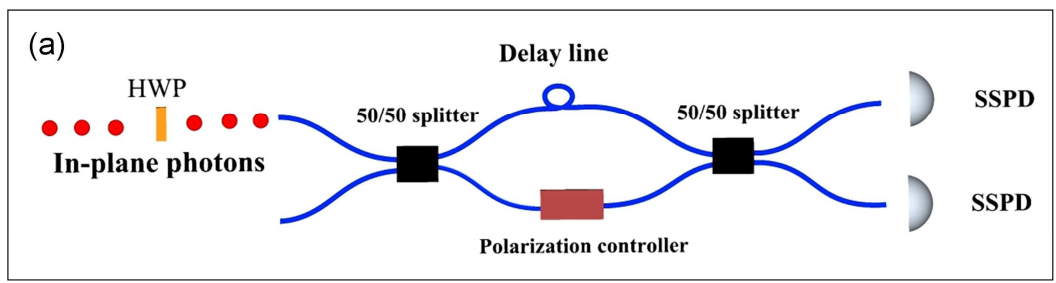

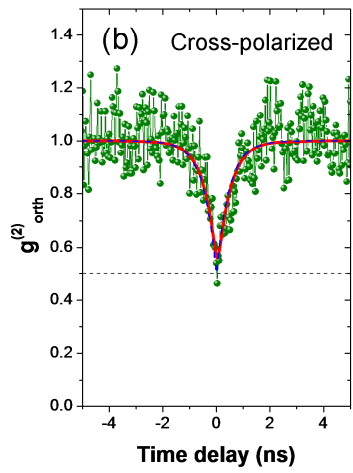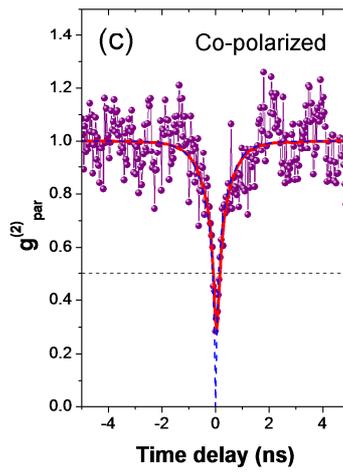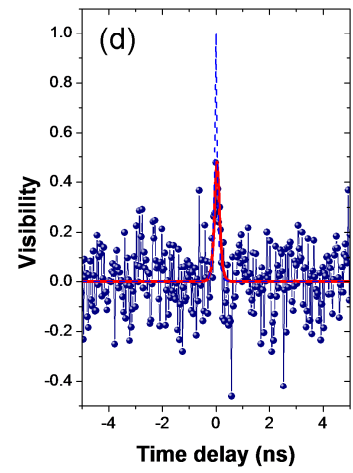